\documentclass[twocolumn,aps,prl]{revtex4}
\usepackage{amssymb}

\usepackage[dvips]{graphicx,graphics}
\usepackage{dcolumn}
\usepackage{bm}

\begin{document}

\title{Direct determination of the crystal field parameters of
Dy, Er and Yb impurities in the skutterudite compound CeFe$_{4}$P$_{12}$
by Electron Spin Resonance.}
\author{D. J. Garcia$^{1,2}$, F. A. Garcia$^{1}$, J. G. S. Duque$^{1}$, P. G. Pagliuso$^{1}$,
C. Rettori$^{1}$, P. Schlottmann$^{3}$, M. S. Torikachvili$^{4}$, and S. B. Oseroff$^{4}$}

\affiliation{$^1$Instituto de F\'isica ``Gleb Wataghin", UNICAMP, Campinas-SP, 13083-970, Brazil.\\
$^2$ Consejo Nacional de Investigaciones Cient\'ificas y T\'ecnicas (CONICET) and Centro At\'omico Bariloche, S.C. de Bariloche, R\'io Negro, Argentina.\\
$^3$Department of Physics, Florida State University, Tallahassee, Florida 32306, U.S.A.\\
$^4$San Diego State University, San Diego, California 92182, U.S.A.}

\date{\today}

\begin{abstract}
Despite extensive research on the skutterudites for the last decade,
their electric crystalline field ground state is still a matter of
controversy. We show that Electron Spin Resonance (ESR)
measurements can determine the full set of crystal field
parameters (CFPs) for the T$_h$ cubic symmetry ($Im3$) of the
Ce$_{1-x}$R$_{x}$Fe$_{4}$P$_{12}$ (R = Dy, Er, Yb, $x\lesssim
0.003$) skutterudite compounds. From the analysis of the ESR data
the three CFPs, $B_{4}^{c}$, $B_{6}^{c}$ and $B_{6}^{t}$ were
determined for each of these rare-earths at the Ce$^{3+}$ site.
The field and temperature dependence of the measured magnetization
for the doped crystals are in excellent agreement with the one
predicted by the CFPs $B_{n}^{m}$ derived from ESR.
\end{abstract}

\maketitle

\section{INTRODUCTION}

The filled skutterudite compounds RT$_{4}$X$_{12}$, where R is a
rare-earth or actinide, T is a transition metal (Fe, Ru, Os) and X
is a pnictogen (P, As, Sb) crystallize in the LaFe$_{4}$P$_{12}$
structure with space group $Im3$ and local point symmetry T$_{h}$
for the R ions. 
Also recently a new skutterudite family, (Sr,Ba)Pt$_4$Ge$_{12}$, was found.\cite{Bauer2}
The R ion is surrounded by eight transition metal
ions forming a cube, and twelve pnictogen ions that form a slightly
deformed icosahedron.\cite{Jeitschko} These materials exhibit a
broad range of strongly correlated electron
phenomena.\cite{Bauer,Dilley,Takeda} In addition, the antimonite
members, are potential thermoelectric materials due to their
enhanced Seebeck coefficient.\cite {Sales,Sekine}

It has been assumed for a long time that the description of the
electric crystalline field (CF) of the cubic point groups, T,
T$_{h}$, O, T$_{d}$, and O$_{h}$ is the same for all of them.
Recently, Takegahara {\it et al.},\cite{Takegahara} studied the
CF for cubic point groups using group theory and a simple point
charge model and found that the above was not correct. Takegahara
{\it et al.} noticed that due to the absence of two symmetry
operations in the T and T$_{h}$ groups, namely the $C_{4}$ and
$C_{2}^{\prime}$ rotations,\cite{Inui} the CF Hamiltonian ($H_{CF}$)
allows for additional sixth order terms with an extra crystal
field parameter (CFP), $B_{6}^{t}$. Therefore, for T$_{h}$
symmetry, in terms of the Steven's operators\cite{Bleaney}
$H_{CF}$ should be written as
\begin{equation}
H_{CF} = B_{4}^{c}(O_{4}^{0} + 5O_{4}^{4}) + B_{6}^{c}(O_{6}^{0} -
21O_{6}^{4}) + B_{6}^{t}(O_{6}^{2} - O_{6}^{6}), \label{HCF}
\end{equation}
where the last term is absent in the ordinary cubic symmetry
O$_{h}$. Its presence does not affect the degeneracy of each
sublevel when compared with that of the O$_{h}$ group, but some
eigenfunctions and eigenvalues may be appreciably
different.\cite{Takegahara} The knowledge of the CF levels,
especially the ground state, is essential to understand the role
of the $4f$-electrons in these compounds. However, in spite
of the large amount of work invested, the CF ground state
is still unclear in several of these systems.\cite{references,Gorem}

Electron spin resonance (ESR) has been used for more than half a
century to examine a wide variety of compounds.\cite{Bleaney} It
is a very useful and highly sensitive technique to study spin
correlations. It provides information about CF effects, site
symmetry, valence of the paramagnetic ions, $g$-value, fine and
hyperfine parameters, etc. Besides, the sample size required for
ESR is typically less than $\sim 4$ mm$^{3}$, i.e., much smaller
than that needed for most other techniques. When the compound
is not paramagnetic, ESR can still provide useful information
by doping the matrix with a small amount of paramagnetic ions
such as Nd$^{3+}$, Gd$^{3+}$, Dy$^{3+}$, Er$^{3+}$, Yb$^{3+}$,
etc. The ESR spectra of the impurities allow not only to learn
about the impurity, but also to study the properties of the host
lattice. In cases where the first excited state is separated
from the ground state by an energy of the order of the
temperature at which the data is taken, a field induced change
of the $g$-value\cite{Oseroff} and an exponentially activated
$T$-dependence of the linewidth\cite{Davidov,Barberis} may be
expected. Moreover, the ESR of an excited state could also
be observed.\cite{Rettori C} Thus, by measuring the ESR at
different frequencies and temperatures of various R impurities,
one may obtain an accurate determination of their ground state
and, in some cases, the full set of CFPs determining the overall
splitting of the ground $J$-multiplet.

Mesquita {\it et al.},\cite{Mesquita} and Martins {\it et al.},\cite{Martins}
measured the ESR spectra of Ce$_{1-x}$R$_{x}$Fe$_{4}$P$_{12}$
(R = Nd, Dy, Er, Yb; $x\lesssim 0.005$) up to 4.2 K. Our data,
taken in the same range of $T$, agree with those published
previously. The data in Refs. \cite{Mesquita,Martins} was
analyzed assuming $H_{CF}$ for the cubic group, i.e., Eq.
(\ref{HCF}) without the $B_6^t$-term. In particular, the
unexpected $g$-value of 6.408(3) measured for the Kramers
doublet ground state of Er$^{3+}$ in CeFe$_{4}$P$_{12}$
cannot be explained if the term $B_{6}^{t}(O_{6}^{2} - O_{6}^{6})$
is not included in $H_{CF}$. By using the $H_{CF}$ given in Eq.
(\ref{HCF}) and measuring up to $T\cong 50$ K to populate the
excited states, the ESR data for the various R impurities can
be explained and the full set of CFPs determined.

The last term in Eq. (\ref{HCF}) is usually of secondary importance.
ESR is the second technique known to us where this term cannot 
be ignored.  The other examples are the crystalline field potential 
of PrOs$_4$Sb$_{12}$ and PrFe$_4$Sb$_{12}$ measured by inelastic 
neutron scattering.\cite{Kuwa,Gorem,Bauer3}. In those compounds the
 $B_6^t$-term rules out the non-Kramers doublet $\Gamma_3$ as the 
ground state, in favor of the $\Gamma_1$ singlet.

\section{EXPERIMENTAL}

Single crystals of Ce$_{1-x}$R$_{x}$Fe$_{4}$P$_{12}$ (R = Nd, Dy,
Er, Yb; $x\lesssim 0.003$) were grown in a molten Sn flux
according to the method described in Ref. \cite{Meisner}.
Whithin the accuracy of microprobe analysis the crystals studied 
are found to be uniform.
The R concentrations were determined from the $H$
and $T$-dependence of the magnetization, $M(H,T)$. $M(H,T)$
measurements were taken in a Quantum Design MPMS SQUID
$dc$-magnetometer. The crystals used were about $2$ x $2$ x $2$
mm$^{3}$ with perfect natural crystallographic grown faces. The
cubic structure (space group $Im3$) and phase purity were checked
by x-ray powder diffraction. The ESR spectra were taken in Bruker
X ($9.48$ GHz) and Q ($34.4$ GHz) band spectrometers using
appropriated resonators coupled to a $T$-controller of a helium
gas flux system for $4.2\lesssim T\lesssim 300$ K. The R$^{3+}$
resonances show dysonian (metallic) lineshape ($A/B\approx 2.5$)
corresponding to a microwave skin depth ($\delta = 1/(\pi
\mu_0 \sigma \nu )^{1/2}$) smaller that the size of the
crystals.\cite{Feher} The low-$T$ metallic character of the
compound is associated to the thermally activated conductivity
($\simeq 10^{-3}$ ($\Omega $cm)$^{-1}$) reported for this material
at low-$T$.\cite{Meisner}

\section{RESULTS}

\begin{figure}[hptb]
\includegraphics[width=\columnwidth]{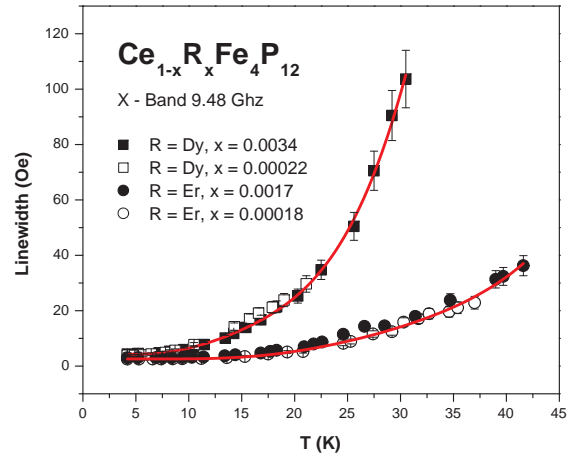}
\caption{(color online) $T$-dependence of $\Delta H$ of
X-band ESR for Dy$^{3+}$ and Er$^{3+}$ in
Ce$_{1-x}$R$_{x}$Fe$_{4}$P$_{12}$ (R = Dy, Er). The solid
lines are fits to Eq. (\ref{DeltaH}) leading to the
following parameters for Dy: $a$ = 4.0(4) Oe, $\Delta_{1}$
= 40(8) K, $\Delta_{2}$ = 135(30) K, $c_{1}$ = 0.0015(2)
Oe/K$^{3}$, $c_{2}$ = 0.0020(2) Oe/K$^{3}$, and for Er:
$a$ = 3.3(3) Oe, $\Delta_{1}$ = 85 (15) K, $\Delta_{2}$ =
300(100) K, $c_{1}$ = 0.0003(1) Oe/K$^{3}$, $c_{2}$ =
0.0002(1) Oe/K$^{3}$.} \label{Fig1}
\end{figure}

\begin{figure}[hptb]
\includegraphics[width=\columnwidth]{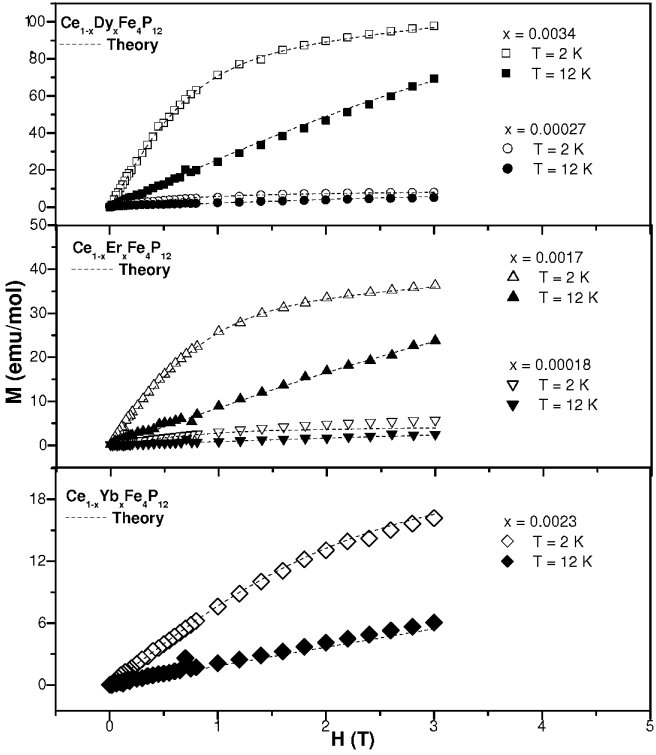}
\caption{$M(H,T)$ for Dy$^{3+}$, Er$^{3+}$ and Yb$^{3+}$ in
Ce$_{1-x}$R$_{x}$Fe$_{4}$P$_{12}$ (R = Dy, Er, Yb). The dashed
curves are the calculated $M(H,T)$ from Eq. (\ref{M}) using
the CFPs from Table I. The contribution of the sample holder
and host lattice to the measured magnetization has been
subtracted.} \label{MxH}
\end{figure}

Fig. \ref{Fig1} shows the $T$-dependence of the X-Band ESR
linewidth, $\Delta H$, for the Kramers doublet ground state of
Dy$^{3+}$ and Er$^{3+}$ in CeFe$_{4}$P$_{12}$. Within the
experimental accuracy, the linear $T$-term is negligible at
low-$T$, in agreement with previous measurements.\cite{Mesquita}
This indicates that there is no spin-lattice relaxation via
an exchange interaction with the conduction-electrons ($ce$)
(Korringa relaxation).\cite{Korringa,Rettori} Q-Band data
(not shown here) are similar to the data presented in Fig.
\ref{Fig1} with slightly larger ($\lesssim 15\%)$ residual
linewidth, $\Delta H$($T$ = 0 K), i.e., no inhomogeneous
broadening is observed. Thus, the exponential increase of
$\Delta H$ at high-$T$ results from an homogeneous line
broadening due to a phonon spin-lattice relaxation process
involving the excited CF levels (see below).\cite{Davidov,Barberis}
For Yb$^{3+}$ in CeFe$_{4}$P$_{12}$ a $T$-independent (not shown)
resonance of $\Delta H$ = 8(2) Oe corresponding to a Kramers
doublet ground state was observed up to $T\simeq 40$ K. Fig.
\ref{MxH} displays $M(H,T)$ for the same samples.

For Nd$^{3+}$ in CeFe$_{4}$P$_{12}$ the ground state corresponds
to an anisotropic quadruplet. The $g$-value anisotropy has been
obtained by measuring the two allowed transitions within this
quadruplet for the field in the $(110)$ plane at $T$ = 4.2 K
\cite{Martins}.

The resonances associated to the above ESR data correspond to the
R$^{3+}$$I=0$ isotopes. We have also observed the resonances
corresponding to various R$^{3+}$ isotopes with $I\neq 0$, that,
at low-$T$, show the same features already
reported.\cite{Mesquita,Martins} Furthermore, the $T$-dependence
of the ESR intensity for the observed resonances follows
approximately a Curie-Weiss law at low-$T$. This indicates that
the resonances arise from the ground state of the CF split
$J$-multiplet. The measured $g$-values and degeneracy of the
ground states are displayed in Table I.

\begin{table*}
\caption{ESR and CFPs for Ce$_{1-x}$R$_x$Fe$_4$P$_{12}$ (R = Nd,
Dy, Er, Yb). Ground state degeneracy is abbreviated ``gsd'' and ``Anis''
denotes anisotropic ground state. 
The symbol ($^{*}$) denotes a result obtained from Dy$^{3+}$ data. \label{TeorTable1}}
\begin{tabular}{|c|c|p{1.4cm}|p{1.4cm}|c|c|c|c|c|c|c|c|c|}
\hline
R$^{3+}$        & gsd & $g_{exp.}$& $g_{calc.}$  &    x  &  y    & $W$[K] & $B_{4}^c$[mK] & $B_{6}^{c}$[mK] & $B_{6}^{t}$[mK] \\
\hline
$Nd^{3+}$ & 4           &Anis.    & Anis.       & -0.566& 0.00  &$<$ 0 &   $>$ 0          &$<$ 0            & 0.00 \\
$Dy^{3+}$ & 2           &7.438(7) & 7.43(3)     &  0.32 & 0.40  & 0.92(16) &   2.9(6)     & 0.027(6)        & 0.28(6) \\
$Er^{3+}$ & 2           &6.408(3) & 6.40(6)     & -0.16 & 0.45  & 1.6(3)   &  -2.3(5)     & 0.053(10)       & 0.54(10) \\
$Yb^{3+}$ & 2           &2.575(2) & 2.6(1)      &  0.54 & 0.08 & 7(2)$^{*}$&   58(17)     & 2.3(7)          & 24(7)  \\
\hline
\end{tabular}
\end{table*}

\section{ANALYSIS AND DISCUSSION}

We now add the Zeeman term $g_{J} \mu _{B}{\bf H} \cdot {\bf J}$ to
Eq. (\ref{HCF}), where $g_{J}$ is the Lande $g$-factor, $\mu_{B}$
the Bohr magneton, ${\bf J}$ the total angular momentum for each R
ion and ${\bf H}$ is the $dc$-magnetic field. Following Lea, Leask
and Wolf (LLW)\cite{Lea} the Hamiltonian can be parametrized as
\begin{eqnarray}
H_{CFZ} & = & W\left\{(1-\vert\mathrm{y}\vert)\left[\mathrm{x}
\frac{O_{4}^c}{F_4^0} + (1-\vert \mathrm{x}\vert)\frac{O_6^{c}}{F_6^0}\right]
+ \mathrm{y} \frac{O_{6}^{t}}{F_6^2} \right\} \nonumber\\
& &+ g_{J} \mu_{B} {\bf H} \cdot {\bf J} , \label{HCFZ}
\end{eqnarray}
where we denoted $O_{4}^{0}+5O_{4}^{4}$ by $O_{4}^c$ and similarly
the sixth order terms by $O_{6}^{c}$ and $O_{6}^{t}$, respectively.
The coefficients of Eq. (\ref{HCF}) are rewritten as $B_4^c =
(1-\vert$y$\vert)$x$W$/F$_4^0$, $B_6^c = (1-\vert$y$\vert)(1-
\vert$x$\vert)W$/F$_6^0$ and $B_6^t = $y$W$/F$_6^2$. The coefficients
F$_n^m$ are tabulated in Ref. \cite{Hutchings} for various
values of $J$. The above is a generalization of the LLW Hamiltonian
that includes the $O_6^t$-term.\cite{Takegahara,Lea} Our
parametrization is slightly different from that in Ref.
\cite{Takegahara} and has the advantage that the entire range of
the CFPs is accounted for within the finite intervals (-1 $\leq $
x $\leq$ 1) and (-1 $\leq$ y $\leq$ 1).

\begin{figure}[htbp]
\begin{center}
\includegraphics[width=\columnwidth]{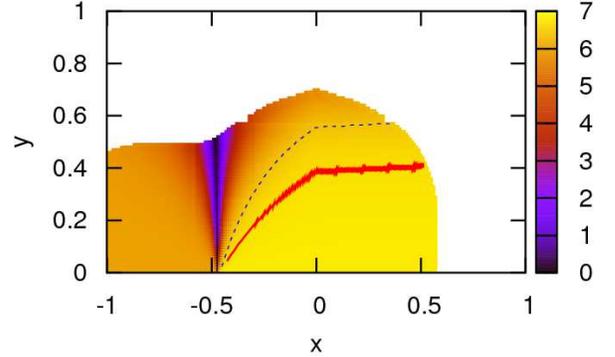}
\end{center}
\caption{(color online)The color scale shows the ground state theoretical $g$-values for 
Er$^{3+}$ ($J=15/2$; $g_J = 6/5$, $W>0$) as a function of (x,y). The blue dashed
line indicates the set of (x,y)-values corresponding to the experimental 
$g = 6.40$. The red line correspond to the (x,y)-values for Dy$^{3+}$ ($J=15/2$;
 $g_J = 4/3$, $W>0$) and measured $g = 7.438$ (the experimental uncertainty 
of the $g$-value is about the width of those lines).\label{Fig2}}
\end{figure}

By diagonalizing $H_{CFZ}$ we obtained the CF wave functions and
energies for each of the R in units of $W$ as a function of
x and y. 
Then, for a small $H$ the doublet ground state ($\Gamma_i$, $i=5,6$ or $7$)
 the g-value can be  calculated 
($g = 2g_J\vert \langle \Gamma_i \vert S_z \vert \Gamma_i \rangle \vert $). 
For finite field and at resonance $g$ can be obtained from the Zeeman 
splitting of the doublet, 
$\Delta E(H)= h\nu = g \mu_B H $.
 Fig. \ref{Fig2} shows the x and y dependence of
the $g$-value for the ground state of Er$^{3+}$ ($J=15/2$;
$g_{J}=6/5$; $W>0$) in a color scale. For y $= 0$ and variable
x, we obtain the expected $g$-value of $6.000$ (orange) for
the $\Gamma_{6}$ and $6.800$ (yellow) for the $\Gamma_{7}$
doublets.\cite{Lea} The white region in Fig. \ref{Fig2}
corresponds to the quadruplet ground states. Thus, a
$g$-value of $6.408$ corresponds neither to a $\Gamma_{6}$
nor to a $\Gamma_{7}$ (y = 0). For y $\neq$ 0 the $g$-value
decreases and approaches to zero for large values of y (black
region). Hence, the measured $g$-value of $\sim 6.4$ for
Er$^{3+}$ corresponds to a doublet ground state with y $\neq$
0. Such a $g$-value is obtained for the set of (x,y) values
indicated by the dashed blue line in Fig. \ref{Fig2}. The
results shown in Fig. \ref{Fig2} do not depend on the sign of y.

The same procedure was followed with the measured $g$-values for
Dy$^{3+}$ ($J=15/2$; $g_{J}=4/3$) and Yb$^{3+}$ ($J=7/2$; $g_{J}
=8/7$) impurities.  To be able to present the  $g$-values of
Dy$^{3+}$ in Fig. 3, we re-scaled them by the g$_J$ ratio between
Dy$^{3+}$ and Er$^{3+}$ (g$^{Dy}_J$/g$^{Er}_J$ = 10/9). The (x,y)
values corresponding to the experimental $g$-value for Dy$^{3+}$
are given by the red curve in Fig. \ref{Fig2}. The results for
Yb$^{3+}$ are similar, but not shown in Fig. \ref{Fig2}.

The $B_n^m$ parameters are angular momentum effective values of
the \textit{actual} CFP $A_n^m$ defined in real space. The $B_n^m$
and $A_n^m$ are related by $B_n^m = \langle r^n \rangle \theta_n
A_n^m$.\cite{Hutchings} Here $\theta_n$ is a geometrical factor
arising from the addition of angular momenta. 
The substitution of a weakly intermediate valence Ce ion by a R$^{3+}$ impurity 
may distort the electron density in the neighborhood of the defect. 
However, the host perturbation by the R$^{3+}$ impurities should be
 comparable for Er$^{3+}$,  Dy$^{3+}$ and Yb$^{3+}$ ions, therefore, 
the \emph{actual} CFPs $A_n^m$ should not depend much on the ion R.
Thus, if R$_1$ and R$_2$ denote two
rare earth impurities, it is possible to relate their CFPs
\cite{Hutchings}
\begin{equation}
\frac{B_n^m ({\rm R}_1)}{\langle r^n({\rm R}_1) \rangle
\theta_n ({\rm R}_1)} = \frac{B_n^m ({\rm R}_2)}{\langle
r^n({\rm R}_2) \rangle \theta_n ({\rm R}_2)}  . \label{Bnm}
\end{equation}
Defining
\begin{eqnarray}
\beta&=&\frac{\langle r^4({\rm R}_2)\rangle}{\langle
r^6({\rm R}_2)\rangle}\frac{\langle r^6({\rm R}_1)\rangle}
{\langle r^4({\rm R}_1)\rangle} \nonumber \\
\delta&=& \frac{\langle r^6({\rm R}_1)\rangle}{\langle
r^6({\rm R}_2)\rangle} \nonumber \\
\xi&=&\frac{\theta_4({\rm R}_2)}{\theta_4({\rm R}_1)}
\frac{\theta_6({\rm R}_1)}{\theta_6({\rm R}_2)}
\frac{F_6^0({\rm R}_1)}{F_6^0({\rm R}_2)}
\frac{F_4^0({\rm R}_2)}{F_4^0({\rm R}_1)} \nonumber \\
\gamma&=&\frac{F_6^0({\rm R}_1)}{F_6^0({\rm R}_2)}
\frac{F_6^2({\rm R}_2)}{F_6^2({\rm R}_1)} \nonumber \\
\eta&=&\frac{F_6^2({\rm R}_1)}{F_6^2({\rm R}_2)}
\frac{\theta_6({\rm R}_1)}{\theta_6({\rm R}_2)}, \nonumber
\end{eqnarray}
we obtain the following relations among the sets of
parameters (x$_2$,y$_2$,$W_2$) and (x$_1$,y$_1$,$W_1$) for
the two ions
\begin{eqnarray}
\mathrm{x}_2 &=& \frac{\xi \beta }{1-(1-\vert \xi \beta \vert)
\vert \mathrm{x}_1\vert} \mathrm{x}_1\nonumber \\
\mathrm{y}_2 &=& \left[1 + \frac{(1-\vert \mathrm{x}_1\vert)
(1-\mathrm{y}_1)}{\gamma \mathrm{y}_1 (1-\vert \frac{\xi \beta
\mathrm{x}_1}{1-(1-\vert \xi \beta \vert)\vert \mathrm{x}_1\vert}
\vert)} \right]^{-1} \label{convxy}
\end{eqnarray}
and
\begin{equation}
W_2 = \eta \delta W_1 \left[\mathrm{y}_1 + \frac{(1-\vert
\mathrm{x}_1\vert) (1-\mathrm{y}_1)}{\gamma(1-\vert
\frac{\xi \beta \mathrm{x}_1}{1-(1-\vert \xi \beta \vert)
\vert \mathrm{x}_1\vert} \vert)} \right]^{-1}\label{convW}
\end{equation}
where $\xi$, $\gamma$ and $\eta$ are \emph{geometrical}
parameters that only depend on $\theta_n$ and $F_n^m$
(their values are tabulated in Ref. \cite{Hutchings}).
On the other hand, $\beta$ and $\delta$ depend on the
expectation values $\langle r^n({\rm R}) \rangle$. $\beta$
enters the expression for x$_2$ and y$_2$, while in order
to obtain $W_2$ also $\delta$ is needed. The values for
$\langle r^n({\rm R}) \rangle$ have been computed in Ref.
\cite{FreeWatson} for the free (unperturbed) rare earth
ions. In general, the $\langle r^n({\rm R})\rangle$ values
depend on the host, in particular wether it is an
insulating \cite{FreemanDeclaux} or a metallic \cite{Lutz}
environment. Their values may be obtained from $ab$-$initio$
calculations, which are beyond of the scope of this work.
Nonetheless, $\beta$ and $\delta$ depend on the $\langle
r^n({\rm R}_1)\rangle$/$\langle r^n({\rm R}_2)\rangle$ ratios that, for
Dy$^{3+}$, Er$^{3+}$ and Yb$^{3+}$, in a given lattice (whether
insulator \cite{FreemanDeclaux,FreeWatson} or metals \cite{Lutz})
present differences smaller than 5\%. In other words, the
changes of $\langle r^n({\rm R}) \rangle$ from the free ion values
are about the same for the various R when located in the same
environment. As CeFe$_4$P$_{12}$ is a small gap semiconductor,
for $\beta$ and $\delta$ we shall assume values close to those
for an insulator. Here we assume that the values are within
$\pm 10\%$ of the insulating ones. \cite{FreemanDeclaux}

\begin{figure}[htbp]
\begin{center}
\includegraphics[width=\columnwidth]{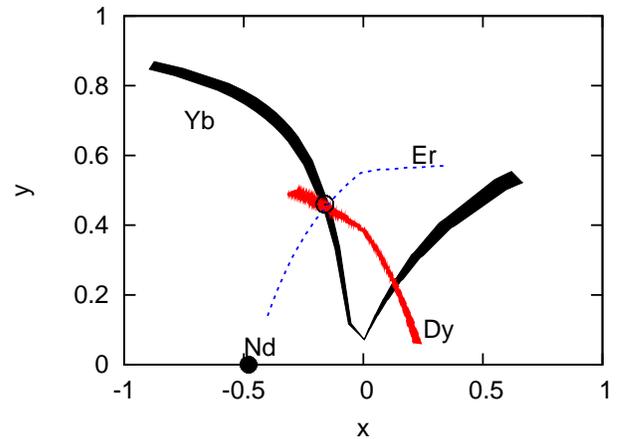}
\end{center}
\caption{(color online) Set of (x,y)-values, satisfying the ground state $g$-values of the studied 
R$^{3+}$ ions, transformed into the Er$^{3+}$ (x,y)-space by eq. 4.
The open circle indicates the point where Dy$^{3+}$, Er$^{3+}$ and
Yb$^{3+}$ share the same ratios, ${A_4^c}\langle r^4({\rm R}) \rangle
/{A_6^c}\langle r^6({\rm R}) \rangle \approx -2.0$ and
${A_6^t}/{A_6^c}\approx 10$ (see text).\label{Fig3}}
\end{figure}

The blue curve in Fig. \ref{Fig3} again shows the (x,y) parameters
for Er$^{3+}$ (see Fig. \ref{Fig2}). Using Eq. \ref{convxy},
the set of (x,y) values for Dy$^{3+}$ and Yb$^{3+}$ that satisfy
the measured ground state $g$-values may be transformed to the
(x,y) space corresponding to Er$^{3+}$. The results for Yb$^{3+}$
and Dy$^{3+}$ are shown in Fig. \ref{Fig3} by the red and black
lines, respectively. The width of these lines includes the
uncertainty of $\beta$ and experimental error bars of the
measured $g$-values. Notice that the lines for Dy$^{3+}$,
Er$^{3+}$ and Yb$^{3+}$ all intersect at a single point
(x $\approx -0.16(3)$, y $\approx 0.45(3)$). 
The three ions have the same charge and a similar size, therefore, we may
 assume that the \textit{actual} CFPs are about the same for these impurities in CeFe$_{4}$P$_{12}$.
This suggests that the
ratios involving the \textit{actual} CFPs are ${A_4^c}\langle
r^4({\rm R}) \rangle /{A_6^c}\langle r^6({\rm R}) \rangle \approx -2.0$ and
${A_6^t}/{A_6^c}\approx 10$ for the three impurities. 
Now these Er$^{3+}$ (x,y)-values are transformed back to obtain the (x,y)-values
 for Dy$^{3+}$ and Yb$^{3+}$, which are listed in Table I. Notice that the (x,y)-values 
for Dy$^{3+}$ and Yb$^{3+}$ are obtained by using their experimental $g$-values and 
the assumption of similar crystal fields.

The above is valid independently of the energy scaling parameter
$W$. The values of $W$ for all three impurities can be determined
if the $W$ for one of them is known (see Eq. (\ref{convW})).
$W$ can be estimated from the $T$-dependence of $\Delta H$ data.
The solid lines seen in Fig. \ref{Fig1} are the best fit of the
measured $\Delta H$ for Dy$^{3+}$ and Er$ ^{3+}$ in CeFe$_{4}$P$_{12}$
to the expression
\begin{equation}
\Delta H = a + c_{1}\frac{\Delta_{1}^{3}}{[e^{\Delta_{1}/kT}-1]}+ c_{2}\frac{
\Delta_{2}^{3}}{[e^{\Delta_{2}/kT}-1]},  \label{DeltaH}
\end{equation}
where $a$ is the residual linewidth. The relaxation is through
phonon modes and requires the coupling of phonons between the
ground and excited CF states. We consider here the two lowest
excited CF states with nonvanishing matrix elements from the
ground state and denote the excitation energies with $\Delta_{1,2}$,
respectively. The coefficients $c_{1,2}$ are given by
$(3k_{B}^{2}/2\pi h^{4}\rho \upsilon^{5}) M_{1,2}^{2}$ ($\rho $ is
the host density, $\upsilon$ the sound velocity and $M_{1,2}^{2}$
the sum of the square of the matrix elements of the
dynamic-crystal field potential). \cite{Davidov,Barberis}
The parameter values resulting from the fits are given in the
caption of Fig. \ref{Fig1}.

\begin{figure}[htbp]
\begin{center}
\includegraphics[width=\columnwidth]{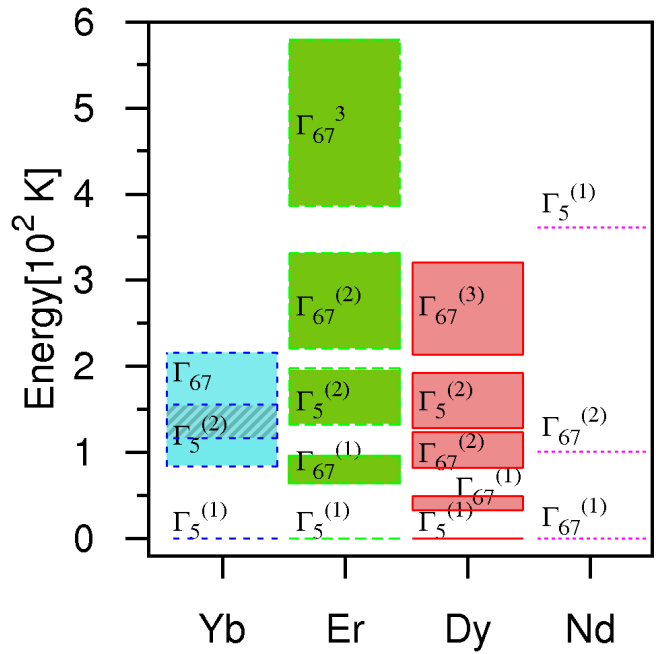}
\end{center}
\caption{(color online) CF energy levels for the studied R$^{3+}$
ions in CeFe$_4$P$_{12}$. For Nd$^{3+}$ we used arbitrarily
$W_{Nd}$ = -5 K. Thus, if the actual $W_{Nd}$ value is known the
Nd$^{3+}$ energy levels should be scaled by -$W_{Nd}/5$ K.
The high of the dotted line boxes indicate the uncertainty of the energy levels.
\label{Niveles}}
\end{figure}

By using the values of (x,y), $\Delta_{1}$ and $\Delta_{2}$
(see Fig. \ref{Fig1}) we obtain $W_{Dy}$ = 0.92(16) K and
$W_{Er}$ = 1.6(3) K. The resulting energy levels for Dy$^{3+}$
and Er$^{3+}$ ions are shown in Fig. \ref{Niveles}. On the
other hand, using $W_{Dy}$ and Eq. \ref{convW} we can obtain
$W$ for the other ions. In particular, for Er$^{3+}$ we obtain
$W_{Er}^{*}$ = 1.3(4) K, where the error bar includes 20\% of
experimental errors and 10\% from the uncertainty of $\delta$.
We see that, within the error bars, the values $W_{Er}$ and
$W_{Er}^{*}$ agree. Using again $W_{Dy}$ we determined
$W_{Yb}^{*}$ = 7(2) K. The Yb$^{3+}$ energy levels are also
shown in Fig. \ref{Niveles}. Once the set of (x,y) and $W$
parameters are known for a given R, their corresponding CFPs
$B_{n}^{m}$ are calculated (see Table I).

Assuming that the $\langle r^n({\rm R})\rangle$ values are, within 10\%,
of their values in insulators, \cite{FreemanDeclaux} the
\textit{actual} CFPs $A_{n}^{m}$ can be estimated
\begin{eqnarray}
&&A_{4}^{c}  \cong -33(10) {\rm K}/a_0^4 , \nonumber \\
&&A_{6}^{c}  \cong 4(1) {\rm K}/a_0^6 , \nonumber \\
&&A_{6}^{t}\cong 44 (15) {\rm K}/a_0^6 , \label{As}
\end{eqnarray}
where $a_0$ is the Bohr radius. The main sources of error
are the experimental $\Delta_1$ and theoretical $\delta$
uncertainties.

For Nd$^{3+}$ the (x,y) values that account for the measured
anisotropic $g$-values of the quadruplet ground state are
given in Table I. These values correspond to a point in the
Er$^{3+}$ (x,y) space (filled circle in Fig. \ref{Fig3}),
which is different than that for the other R$^{3+}$ ions.
This suggests that the large Nd$^{3+}$ ionic radius, as
compared with those of the other R$^{3+}$ and the
intermediate valence of the Ce ions, probably causes a
large local crystal distortion close to the Nd$^{3+}$ site.
Additional experimental information, involving the excited
CF levels, would be needed  to determine $W_{Nd}$ in this
compound. Therefore, the complete set of \textit{actual}
CFPs $A_{n}^{m}$ for Nd$^{3+}$ cannot be given. Nevertheless,
the (x,y) values for Nd$^{3+}$ are compatible with $W_{Nd} < 0$.
In Fig. \ref{Niveles} we present the Nd$^{3+}$ energy levels
using arbitrarily $W_{Nd}$ = -5 K (see caption of Fig. \ref{Niveles}).

The CF splittings of the $J$-multiplet determine $M(H,T)$
\begin{equation}
M(H,T)=\frac{\sum_{i=1}^{2J+1} m_i(H) e^{-E_i(H)/k_B
T}}{\sum_{i=1}^{2J+1} e^{-E_i(H)/k_B T}}, \label{M}
\end{equation}
where $m_i(H)$ and $E_i(H)$ are the magnetization and energy
eigenvalue of each eigenstate of the Hamiltonian (Eq. \ref{HCFZ})
computed at a \textit{finite} $H$ using Eq. \ref{M} and the
CFPs $B_{n}^{m}$ given in Table I. The dashed curves in Fig.
\ref{MxH} show the calculated magnetization, $M(H,2$ K) and
$M(H,12$ K), for two concentrations of Dy$^{3+}$, Er$^{3+}$
and Yb$^{3+}$ as compared to the experimental data. 
In all cases the sample holder diamagnetism was previously determined 
and subtracted from the total magnetization. 
Also, the paramagnetic contribution of the undoped CeFe$_4$P$_{12}$ host
lattice was measured and subtracted. 
At low-T ($2$ K $\leqslant$ T $\leqslant 12$ K) that magnetization is less 
than 10 \% of the samples doped with Er ($x=0.0017$) and Dy ($x=0.0034$) and
 $\sim 30$ \% of the one doped with Yb ($x=0.0023$).

\section{CONCLUSIONS}

The filled skutterudite CeFe$_{4}$P$_{12}$ compound is a
small gap ($\simeq 1500$ K) semiconductor.\cite{Meisner} Hence,
the R$^{3+}$ spin-lattice relaxation via an exchange interaction
with $ce$ is inhibited (Korringa process),\cite{Korringa,Rettori}
since the $ce$ must be promoted via exponential activation.
This is verified by the absence of a linear $T$-term in our
low-$T$ $\Delta H$ data (see Fig. \ref{Fig1}). Similarly, a
$g$-shift (Knight shift) \cite{Rettori} is not expected.
Therefore, the shift of the $g$-value of the Kramers
ground-doublet relative to that in O$_h$ symmetry (y = 0)
is due to the $B_{6}^{t} (O_{6}^{2} - O_{6}^{6})$-term in
$H_{CF}$. For O$_h$ symmetry the Kramers doublet $g$-values
are unique (independent of the CFPs) \cite{Lea} and the
exchange coupling in a metallic host is simple obtained
from the $g$-shift of the resonance. Our calculation showed
that the presence of the new term results always in an
isotropic $g$-value and a negative $g$-shift for the doublet
ground states for $J$ = 7/2 and 15/2. For impurities in
metallic hosts with T$_h$ symmetry, when studied by ESR,
a negative $g$-shift results in a complication to evaluate
the sign and magnitude of the exchange interaction between
the R$^{3+}$ localized magnetic moment and the $ce$.

In summary, in this work we measured the ESR for Dy$^{3+}$,
Er$^{3+}$ and Yb$^{3+}$ ions doped into the filled skutterudite
CeFe$_{4}$P$_{12}$ with T$_h$ structure. We obtained the three
CFPs $B_{n}^{m}$, determined the CF ground state, explained
the unexpected Er$^{3+}$ $g$-value, and found the CF overall
splitting for the $J$-ground state multiplet. With the obtained
CFPs we could fit the low-$T$ $M(H,T)$ of the crystals used in
the ESR experiments.
Moreover, our working assumption that the \textit{actual} CFPs $A_{n}^{m}$
 are about the same  for Dy$^{3+}$, Er$^{3+}$ and Yb$^{3+}$ in this 
compound turned out to be very plausible.
A similar work could be carried out on
undoped compounds such as LnFe$_{4}$P$_{12}$, for Ln = Nd, Gd,
Dy, etc., all Kramers ions with magnetic ground multiplet.
Besides, this work and our preliminary ESR data in the doped
unfilled skutterudites CoSb$_{3}$ put in evidence the importance
of the extra $B_{6}^{t}(O_{6}^{2} - O_{6}^{6})$-term in $H_{CF}$
for compounds with T$_h$ symmetry. In addition, we emphasized
the extra caution we need to have when ESR is used to determine
the exchange parameter in metallic compounds with T and T$_h$
symmetry.

\section*{ACKNOWLEDGMENTS}
The work at UNICAMP was supported by FAPESP and CNPq, Brazil.
P.S is supported by the U.S. Department of Energy via grant No.
DE-FG02-98ER45707.

\end{document}